\documentclass{tlp}
\usepackage{times}
\usepackage{helvet}
\usepackage{courier}
\usepackage{amsmath}
\usepackage{graphicx}
\usepackage{xspace}
\usepackage[defblank]{paralist}
\usepackage{latexsym}

\newcommand{\nesubset}{\rotatebox[origin=c]{45}{$\mathbf{\subset}$}}
\newcommand{\sesubset}{\rotatebox[origin=c]{-45}{$\mathbf{\subset}$}}

\title{Annotated Defeasible Logic}

\author[G. Governatori \and M.J. Maher]
{Guido Governatori \\
Data61, CSIRO, Australia    \\
E-mail: guido.governatori@data61.csiro.au
\and
Michael J. Maher \\
Reasoning Research Institute, Australia    \\
E-mail: michael.maher@reasoning.org.au
}

\newcommand{\ignore}[1]{}
\newcommand{\finish}[1]{}
\newcommand{\skipit}[1]{{ #1}}

\newcommand{\DL}{{\bf DL}}

\newcommand{\mt}[1]{\mathtt{#1}}
\newcommand{\supp}{\sigma}   % support

\newcommand{\cC}{{\cal C}}
\newcommand{\cM}{{\cal M}}
\newcommand{\cS}{{\cal S}}

\newcommand{\non}{{\sim\!\:\!}}

\newcommand{\mi}[1]{\mathit{#1}}

\newcommand{\seq}[2][n]{\ensuremath{#2_{1},\dots,#2_{#1}}\xspace}

\newcounter{fact}
\def\thefact{}
{\end{tabbing}}
\newenvironment{Fact}{\refstepcounter{fact}%
\begin{tabbing}
cxxxx\=xxx\=xxx\=\kill
\thefact\>\+}%
{\end{tabbing}}

\newcounter{clause}
\def\theclause{$c$\arabic{clause}}

\newenvironment{clause}{\begin{tabbing}
xxx\=xxx\=xxx\=\+\kill}%
{\end{tabbing}}

\newenvironment{Clause}{\refstepcounter{clause}%
\begin{tabbing}
cxxxx\=xxx\=xxx\=\kill
\theclause\>\+}%
{\end{tabbing}}

\newcommand{\fail}{\mt{fail}}
\newcommand{\free}{\mt{free}}

\newtheorem{theorem}{Theorem}

\newtheorem{propn}[theorem]{Proposition}
\newtheorem{corollary}[theorem]{Corollary}
\newtheorem{example}[theorem]{Example}
\newtheorem{conjecture}[theorem]{Conjecture}

\begin{document}

\maketitle

\label{firstpage}

\begin{abstract}
Defeasible logics provide several linguistic features to support the expression of defeasible knowledge.
There is also a wide variety of such logics, expressing different intuitions about defeasible reasoning.
However, the logics can only combine in trivial ways.
This limits their usefulness in contexts where different intuitions are at play
in different aspects of a problem.
In particular, in some legal settings, different actors have different burdens of proof,
which might be expressed as reasoning in different defeasible logics.

In this paper,
we introduce annotated defeasible logic as a flexible formalism permitting multiple forms of defeasibility,
and establish some properties of the formalism.

This paper is under consideration for acceptance in {\em Theory and Practice of Logic Programming}.
\end{abstract}
\begin{keywords}
defeasible logic, non-monotonic reasoning, annotated logics, legal reasoning
\end{keywords}

\section{Introduction}

In some application domains, for example legal reasoning, knowing that something
holds (or it is presumed to hold) is not enough to draw further conclusions from it. 
One has to determine to what degree one can assert that it holds. In other words
statements in rules (here we use the term `rule' to indicate a mechanism/principle
to assert conclusions from already established assertions) have an associated proof
standard. Accordingly, a party wanting to assert a particular assertion has the burden
to prove that assertion with the appropriate standard (or a stronger one).  Consider the 
following rule:
\[
  \mathit{IllegalBehaviour}, \neg\mathit{Justification} \Rightarrow \mathit{Liability}
\]
Suppose there is factual evidence about the illegal behaviour. The information in the rule
is not enough, since it does not prescribe the burden needed to assess whether the behaviour 
was justified or not. According to \cite{PrakkenSartor2007,jurix10burden}, in a civil case the lack of 
justification is subject to the so-called \emph{burden of production}, i.e., there is a 
credible argument for it, while in a criminal case the \emph{burden of persuasion} applies (i.e., 
more sceptical reasoning must be used).

Let us consider a concrete scenario.  Party A 
caused some injuries to B. 
Party A was much stronger than Party B, and thus the action causing injury is not justified.
On the other hand,  Party A claims that they acted in self defence since they were 
under threat from Party B. The scenario can now be modelled by the rules:
\[
\begin{array}{lrclclrcl}
& \mi{Injury}, \neg \mi{Justification} & \Rightarrow & \phantom{\neg} \mi{Liability}\\
& \mi{Threat} & \Rightarrow & \phantom{\neg} \mi{Justification}\\
& \mi{Stronger} & \Rightarrow &\neg \mi{Justification} \\
\end{array}
\]
Here, in case we are not able to assess whether the threat was real, 
%despite Party A being a dominant position, 
we have a credible argument for $\neg\mi{Justification}$ (because Party A is stronger), but 
we do not have a sceptical argument for it (because it might be that the threat was real, and then the outcome
from the two conflicting rules is undetermined). 
Thus, we can establish liability in a civil case, 
but Party A is not criminally liable.  Accordingly, we can reformulate the initial rule
in the following two principles: 
\[
\begin{array}{lrclclrcl}
&  \mi{Tort}, \mi{BurdenProduction}(\neg \mi{Justification}) & \Rightarrow & \mi{CivilCaseLiability}\\
&  \mi{Crime}, \mi{BurdenPersuasion}(\neg \mi{Justification}) & \Rightarrow & \mi{CriminalCaseLiability} \\
\end{array}
\]
where $\mi{BurdenProduction}$ and $\mi{BurdenPersuasion}$ are annotations describing the mode 
in which we have to prove the lack of justification for the illegal behaviour. 

Legal reasoning has developed so-called \emph{proof standards} (e.g., scintilla of evidence, 
substantial evidence, preponderance of evidence, beyond reasonable doubt) according to which
assertions have to be justified. \cite{GordonWalton:proof} proposed to encode proof standards
using rule-based argumentation with salience, and \cite{icail2011carneades} shows how to 
represent the proof standards of \cite{GordonWalton:proof} where, essentially, each 
proof standard corresponds to a different degree of provability in some defeasible logic 
variant. In particular,
\cite{icail2011carneades} argues that the proof standard of beyond reasonable doubt 
corresponds to provability in the ambiguity propagating variant of defeasible logic.
% while the preponderance of evidence proof standard corresponds to provability in
% the ambiguity blocking variant.
However, as the following example illustrates there are examples where more than one such proof 
standards must be used. This means that incompatible variants of defeasible 
logic have to work side-by-side. 

Suppose that a piece of evidence $A$ suggests that the defendant in a legal case 
is not responsible while a second piece of evidence $B$ indicates that he/she is
responsible; moreover, the sources are equally reliable.  According to the underlying 
legal system a defendant is presumed innocent (i.e., not guilty) unless responsibility 
has been proved (beyond reasonable doubt).

The above scenario is encoded by the following rules:
\[
\begin{array}{lrclclrcl}
r_1: & \mathit{EvidenceA} &\Rightarrow & \neg\mathit{Responsible}\\
r_2: & \mathit{EvidenceB} &\Rightarrow  & \phantom{\neg} \mathit{Responsible}& ~~~~~ &
r_3: & \mathit{Responsible} &\Rightarrow &  \phantom{\neg} \mathit{Guilty}\\
&&&&&  r_4: &{} & \Rightarrow & \neg \mathit{Guilty}
\end{array}
\]
where $r_{3}$ is stronger than $r_{4}$.
Given both $\mathit{EvidenceA}$ and $\mathit{EvidenceB}$, the literal
$\mathit{Responsible}$ is ambiguous. There are applicable rules ($r_1$ and
$r_2$) for and against the literal, with no way to adjudicate between them. As a
consequence $r_3$ is not applicable, and so there is no applicable rule
arguing against the presumption of innocence (rule $r_{4}$).
In an ambiguity blocking setting we obtain a $\neg\mathit{Guilty}$ verdict;
the ambiguity about responsibility is blocked from applying to $\mathit{Guilty}$. 
In contrast, in an ambiguity propagating setting, 
the ambiguity of $\mi{Responsible}$ propagates to $\mathit{Guilty}$, and thus the literals
$\mathit{Guilty}$ and $\neg \mathit{Guilty}$ are ambiguous too; hence an undisputed
conclusion cannot be drawn.
When we look at the example above, is it appropriate to say that we have reached
a not guilty verdict without any reasonable doubt? 
The evidence supporting that the defendant was responsible has not been refuted.
This example supports the contention of \cite{icail2011carneades} that
ambiguity propagating inference is a more appropriate representation of
proof beyond a reasonable doubt.

Let us extend the scenario. Suppose that the legal system allows for compensation 
for wrongly accused people. A person (defendant) has been wrongly accused if 
the defendant is found innocent, where innocent is defined as $\neg\mi{Guilty}$. 
In addition, by default, people are not entitled to compensation. The additional 
elements of this scenario are modelled by the rules:
\[
\begin{array}{lrclclrcl}
    r_5:& \neg\mi{Guilty} & \Rightarrow &\phantom{\neg}  \mi{Compensation}\\
    r_6:& {} & \Rightarrow & \neg\mi{Compensation} \\
\end{array}
\]
where $r_5$ is stronger than $r_6$.

In the full scenario, the defendant is not found innocent, and so is not entitled to compensation.

If we take a purely ambiguity blocking stance then, since we are not able to determine 
whether there was responsibility, the defendant is not guilty, and then the 
defendant is entitled to compensation.  On the other hand, in a purely ambiguity 
propagating setting, $\mi{Guilty}$ and $\neg\mi{Guilty}$ are ambiguous, 
and this makes $\mi{Compensation}$ and $\neg\mi{Compensation}$ ambiguous; 
we are in a position where we cannot decide whether the defendant is 
entitled or not to compensation.
Thus, both choices are unsatisfactory:
either the defendant receives compensation despite not being found innocent
or no decision is made about compensation.

What we want is a regime where we can reason about guilt in an ambiguity propagating way,
but then reason about compensation in an ambiguity blocking way.
This can be achieve by replacing rule $r_{5}$ with 
\[
 r'_{5}: \mi{BeyondReasonableDoubt}(\neg\mi{Guitly}) \Rightarrow \mi{Compensation}
\] 
where, similarly to what we have done in the previous example, $BeyondReasonableDoubt$
is an annotation to the literal $\neg\mi{Guilty}$ that holds in case the literal is
provable under ambiguity propagation,
and the proof standard for $\mi{Compensation}$ can be chosen to be ambiguity blocking.

The purpose of this paper is to provide a formalism -- \emph{annotated defeasible logic} --
in which such distinctions can be expressed,
define its semantics, and investigate properties of the formalism.

This paper is organised as follows.
In the next section we provide brief background on defeasible logics.
We then introduce annotated defeasible logic,
and define its behaviour with a meta-program.
In the following section we establish some properties of annotated defeasible logic,
including its relationship to existing defeasible logics and
the relative inference strength of the additional inference rules we introduce.
Finally, we show that annotated defeasible logic has the flexibility to deal with different
notions of failure, corresponding to different semantics of negation-as-failure in logic programs.
Due to space limitations, parts of the paper -- including proof sketches -- are presented
in the supplementary material accompanying the paper at the TPLP archive.

\section{Defeasible Logics} 

In this section we can only present an outline of defeasible logics.
Further details can be obtained from \cite{TOCL10} and the references therein.
We address propositional defeasible logics,
but many results should extend to a first-order language.

A defeasible theory is built from a language $\Sigma$ of literals (which we assume is closed under negation)
and a language $\Lambda$ of labels.
A \emph{defeasible theory} $D = (F, R, >)$ consists of a set of facts $F$, a finite set of rules $R$,
each rule with a distinct label from $\Lambda$,
and an acyclic relation $>$ on $\Lambda$ called the \emph{superiority relation}.
This syntax is uniform for all the logics considered here.
Facts are individual literals expressing indisputable truths.
Rules relate a set of literals (the body), via an arrow, to a literal (the head), and are one of three types:
a strict rule, with arrow $\rightarrow$;
a defeasible rule, with arrow $\Rightarrow$;
or
a defeater,  with arrow $\leadsto$.
Strict rules represent inferences that are unequivocally sound if based on definite knowledge;
defeasible rules represent inferences that are generally sound.
Inferences suggested by a defeasible rule may fail, due to the presence in the theory
of other rules.
Defeaters do not support inferences, but may impede inferences suggested by other rules.
The superiority relation provides a local priority on rules with conflicting heads.
Strict or defeasible rules whose bodies are established defeasibly represent claims
for the head of the rule to be concluded.
When both a literal and its negation are claimed,
the superiority relation contributes to the adjudication of these conflicting claims by an inference rule,
leading (possibly) to a conclusion.

Defeasible logics derive conclusions that are outside the syntax of the theories.
Conclusions may have the form 
${+}d q$, which denotes that under the inference rule $d$ the literal $q$ can be concluded,
or
$-d q$, which denotes that the logic can establish that under the inference rule $d$ the literal $q$ cannot be concluded.
The syntactic element $d$ is called a proof tag.
In general, neither conclusion may be derivable:
$q$ cannot be concluded under $d$, but the logic is unable to establish that.
Tags ${+}\Delta$ and $-\Delta$ represent monotonic provability (and unprovability)
where inference is based on facts, strict rules, and modus ponens.
We assume these tags and their inference rules are present in every defeasible logic.
What distinguishes a logic is the inference rules for defeasible reasoning.
The four logics discussed in \cite{TOCL10} correspond to four different pairs of inference rules,
tagged $\partial$, $\delta$, $\partial^*$, and $\delta^*$;
they produce conclusions of the form (respectively) ${+}\partial q$, $-\partial q$, ${+}\delta q$, $-\delta q$, etc.,
where $q$ is a literal. 
These logics all abide by the Principle of Strong Negation \cite{flexf}, which asserts that the condition
for applying a $-d$ inference rule should be the strong negation of the condition for applying ${+}d$.
The inference rules $\delta$ and $\delta^*$ require auxiliary tags and inference rules,
denoted by $\supp_\delta$ and $\supp_{\delta^*}$, respectively\footnote{
Note that in previous works these have been denoted by $\supp$ and $\supp^*$ or $\int$ and $\int^*$.
This change of notation is made to accommodate new forms of support introduced in this paper.
},
expressing that there is at least (weak) support for the conclusion.
These inference rules are available in the supplementary material.
For each of the four principal defeasible tags $d$, the corresponding logic is denoted by $\DL(d)$.
We write $D \vdash +d q$ (respectively, $D \vdash -d q$) if $+d q$ ($-d q$) can be proved by $\DL(d)$.

The four principal tags and corresponding inference rules represent different intuitions about defeasible reasoning,
that is, define different forms of defeasibility:
in $\partial$ and $\partial^*$ ambiguity is blocked, while in $\delta$ and $\delta^*$ ambiguity is propagated;
in $\partial$ and $\delta$ rules for a literal act as a team to overcome competing rules, 
while in $\partial^*$ and $\delta^*$ an individual rule must overcome all competing rules.
The scenario in the introduction with rules $r_1 - r_4$ exemplifies the treatments of ambiguity.
For an example of team defeat, consider rules $s_1$ and $s_2$ for $q$ and rules $s_3$ and $s_4$ for $\neg q$,
with $s_1 > s_3$ and $s_2 > s_4$;
then no individual rule for $q$ can overcome the rules for $\neg q$,
but $s_1$ and $s_2$ -- as a team -- can,
because every rule for $\neg q$ is overridden by some rule in the team.
A more detailed discussion of ambiguity and team defeat in the $\DL$ framework is given in \cite{TOCL10,Maher12}.

In \cite{MG99,flexf}, the inference rules in $\DL(d)$ were reformulated as a meta-program $\cM_d$:
a logic program that takes a representation of a defeasible theory $D$ as input and specifies
what conclusions can be drawn from the theory according to the $d$ inference rules.
(The combined meta-program and theory is denoted by $\cM_d(D)$.)
We will take this meta-programming formulation as our starting point, 
rather than the inference rules as presented in \cite{TOCL10},
for example.
This meta-program formulation is given in the supplementary material.
We assume, initially, that the logic programming semantics in use is
Kunen's semantics \cite{Kunen}, which expresses the 3-valued logical consequences of
the Clark completion of a logic program.
Equivalently, 
Kunen's semantics is the set of all consequences of $\Phi \uparrow n$ for any finite $n$,
where $\Phi$ is Fitting's semantic function for logic programs \cite{Fitting}.
(Fitting's semantics, which is the least fixedpoint of $\Phi$, expresses the 
logical consequences of 3-valued Herbrand models of
the Clark completion of a logic program.)

Although defeasible logics are usually founded on proofs, there are alternative semantics for these logics:
a model-theoretic semantics was defined in \cite{Maher02}, 
a denotational semantics for $\DL(\partial)$ was presented in \cite{densem},
and an argumentation semantics for $\DL(\partial)$ was given in \cite{JLC04}.
Each of these approaches provides an alternative characterization of the
conclusions derivable by proofs in the logic.
However, in this paper we only use the meta-programming formulation of the proof systems.

In the following, annotated defeasible logic will be defined as an integration of the four defeasible logics discussed above.
However, it should be clear that the same approach can be applied to any set of defeasible logics
employing the same logic programming semantics.

\section{Annotated Defeasible Logic}   \label{sect:ADL}

Annotated defeasible logic is the formalism we propose, motivated by the discussion in the introduction.
We begin by addressing its syntax, which is an extension of the syntax of defeasible logics.

A tag is any one of the proof tags, or the additional tag $\free$.
An annotated literal has the form $t \: q$, where $t$ is a tag and $q$ is a literal.
An \emph{annotated defeasible rule} has the form
\[
r: ~~ L_1, \ldots, L_n  \Rightarrow q
\]
where $r$ is a label, $q$ is a literal and each $L_i$ is either an annotated literal or a fail-expression,
where a \emph{fail-expression} has the form $\fail ~ L$, where $L$ is an annotated literal.
An annotated defeater is defined similarly; strict rules are not annotated.

Roughly, the meaning of a rule
\[
r: ~~ t_1 \: q_1, \ldots, t _n\: q_n,  \fail\: t _{n+1}\: q_{n+1}, \ldots, \fail\: t_m \: q_m \Rightarrow q
\]
is that if $q_i$ can be proved using inference rule $t_i$, for $1 \leq i \leq n$,
and proof of $q_i$ can be demonstrated to fail using inference rule $t_i$, for $n+1 \leq i \leq m$,
then we have a \emph{prima facie} reason to infer $q$.
As with all defeasible logics, such an inference can be overridden by another rule.

A proof tag only indicates which inference rule should be applied to resolve conflict concerning that literal.
Thus, an annotated literal $t \: q$ is asking, roughly, for ${+}t \: q$ to be proved.
A fail-expression $\fail \: t \: q$ is asking, roughly, for ${-}t \: q$ to be proved.
The $\free$ tag has a different meaning than the proof tags.
A free literal $\free q_i$ must be proved by the same inference rule that is intended to prove $q$.
This provides a mechanism by which defeasible rules can be agnostic as to inference rule,
which can be determined later,
just as defeasible rules in current defeasible logics are.

An \emph{annotated defeasible theory} is a defeasible theory where the defeasible rules are annotated
and fail-expressions are allowed.
Alternatively, we can think of an annotated defeasible theory as consisting of 
an unannotated defeasible theory (the  \emph{underlying theory}) $D$ that allows fail-expressions, and 
an \emph{annotation function} $\alpha$ that maps each body literal occurrence to its annotation.
In this case we denote the annotated defeasible theory by $\alpha(D)$.
We can consider $\alpha$ a total function, or consider it a partial function mapping literal occurrences to proof tags.
The unmapped  literals are $\free$.

We now turn to expressing the meaning of annotated defeasible theories using the meta-programming approach.
The semantics of a theory is parameterized by a logic programming semantics, which is applied to a meta-program.

Given an annotated defeasible theory $D=(F,R,>)$, the theory is represented by facts as follows:
%\newpage
\begin{enumerate}
\item $\mt{fact}(p)$.  \hfill if $p\in F$
\item $\mt{strict}(r_i,p,[L_1,\dots,L_n])$. 
\hspace*{\fill}
if
  $r_i:L_1,\dots,L_n\to p\in R$
\item $\mt{defeasible}(r_i,p,[L_1,\dots,L_n])$.
\hspace*{\fill}
if
  $r_i:L_1,\dots,L_n\Rightarrow p\in R$
\item $\mt{defeater}(r_i,p,[L_1,\dots,L_n])$. 
\hspace*{\fill}
if 
  $r_i:L_1,\dots,L_n\leadsto p\in R$
\item $\mt{sup}(r_i,r_j)$. 
\hspace*{\fill}
for each pair of rules such that
  $r_i > r_j$
\end{enumerate}
where the $L_i$ are annotated literals or fail-expressions.

The meta-program to which these facts are input is denoted by $\cM$, while
the combination of $\cM$ and the representation of $D$ is denoted by $\cM(D)$.
In what follows, we permit ourselves some syntactic flexibility in presenting
the meta-program.  (For example, we enumerate a list instead of explicitly iterating over it, 
and express the complementation operation $\non$ as a function\footnote{
The complement of $p$ is $\neg p$ and the complement of $\neg p$ is $p$.  
$\non$ is unrelated to $\fail$, since it is the complement of classical negation.}.
Furthermore, tags and $\fail$ are unary functors.)
However, there is no technical difficulty in using
conventional logic programming syntax to represent this program.

Before we get to the predicates that define the meaning of theories, we define some auxiliary predicates.

As discussed in the introduction to defeasible logics,
the different proof tags represent different forms of defeasibility.
In particular, some forms block ambiguity, while others propagate ambiguity;
some use team defeat, while others require an individual rule to overcome all conflicting rules.
The following facts are used to specify, for each proof tag:
that it is a proof tag, 
whether it expresses team defeat or individual defeat,
and whether the inference rule blocks or propagates ambiguity.
Strictly speaking, we should distinguish the proof tags appearing syntactically in $\cM$
from the tags appearing in conclusions (which are not part of the syntax of defeasible logics, but part of its meta-theory).
However, because there is a clear correspondence between the two, we find it clearer to use the same symbol for both. \\

{\small
\begin{minipage}{0.3\textwidth}
\begin{Fact}\label{f:teamB}
$\mt{team}( \partial ).$
\end{Fact}
\begin{Fact}\label{f:teamP}
$\mt{team}( \delta ).$
\end{Fact}
\begin{Fact}\label{f:indivB}
$\mt{indiv}( \partial^* ).$
\end{Fact}
\begin{Fact}\label{f:indivP}
$\mt{indiv}( \delta^* ).$
\end{Fact}
\end{minipage} \hfill
\begin{minipage}{0.3\textwidth}
\begin{Fact}\label{f:Iblock}
$\mt{ambiguity\_blocking}( \partial^* ).$
\end{Fact}
\begin{Fact}\label{f:Tblock}
$\mt{ambiguity\_blocking}( \partial ).$
\end{Fact}
\begin{Fact}\label{f:Iprop}
$\mt{ambiguity\_propagating}( \delta^* ).$
\end{Fact}
\begin{Fact}\label{f:Tprop}
$\mt{ambiguity\_propagating}( \delta ).$
\end{Fact}
\end{minipage} \hfill
\begin{minipage}{0.3\textwidth}
\begin{Fact}\label{f:tagTB}
$\mt{proof\_tag}( \partial^* ).$
\end{Fact}
\begin{Fact}\label{f:tagIB}
$\mt{proof\_tag}( \partial ).$
\end{Fact}
\begin{Fact}\label{f:tagTP}
$\mt{proof\_tag}( \delta^* ).$
\end{Fact}
\begin{Fact}\label{f:tagIP}
$\mt{proof\_tag}( \delta ).$
\end{Fact}
\end{minipage}
} \\\ \\

The following clauses define the class of all rules and the class of supportive rules.
Defeaters are not supportive rules because they can only be used to prevent other conclusions;
they cannot support any conclusion.

{\small
\begin{clause}
$\mt{supportive\_rule}(Label,Head,Body)$:-\\
  \> $\mt{strict}(Label,Head,Body)$.\\
\\[-.5\baselineskip]
$\mt{supportive\_rule}(Label,Head,Body)$:-\\
  \> $\mt{defeasible}(Label,Head,Body)$.
\end{clause}

\begin{clause}
  $\mt{rule}(Label,Head,Body)$:-\\
  \> $\mt{supportive\_rule}(Label,Head,Body)$.\\
\\[-.5\baselineskip]
  $\mt{rule}(Label,Head,Body)$:-\\
  \> $\mt{defeater}(Label,Head,Body)$.  
\end{clause}
}

The next clauses express monotonic provability. \\

{\small
\begin{Clause}\label{strictly1}
  $\mt{definitely}(X)$ :-\\
  \> $\mt{fact}(X)$.
\end{Clause}

\begin{Clause}\label{strictly2}
  $\mt{definitely}(X)$ :-\\
  \> $\mt{strict}(R,X,[\seq{Y}])$,\\
  \> $\mt{definitely}(Y_1)$,\dots,$\mt{definitely}(Y_n)$.
\end{Clause}
}

In the predicate expressing defeasible inference, $\mt{defeasibly}$,
one argument is written as a subscript $\mt{Z}$ in the following clauses.
That argument takes as its value one of the four proof tags and represents
the inference rule that should be applied to resolve conflict for the literal in the other argument,
unless the literal has a proof annotation.
All clauses for predicates with a subscript $\mt{Z}$
implicitly contain $\mt{proof\_tag}(\mt{Z})$ in their body.
In clause \ref{defeasibly_free} we see that $\free$-annotated literals are to be proved according to $\mt{Z}$.

In clause \ref{defeasibly_fail}, fail-expressions are defined: failure is implemented by negation.
This is valid because the logics involved satisfy the Principle of Strong Negation.
For such logics, the conditions for $-d$ inference rules are a negation of the conditions for $+d$ inference rules.
In both defeasible logics and logic programming, failure-to-prove is a primitive notion,
available in defeasible logics through negative tags and in logic programming through negation.
Hence, it is not surprising that failure is implemented by negation in the meta-program.

The remaining two clauses are reflective of the basic structure of defeasible reasoning.
Clause \ref{defeasibly1} expresses that any literally that is definitely true 
(proved monotonically from facts and strict rules)
is also defeasibly true.
Clause \ref{defeasibly_notfree} handles an annotated literal by using the tag $Y$
as the subscript argument in subsidiary computations.
This clause says that a literal $X$, annotated by $Y$, is proved if
the negation of $X$ is not proved monotonically
and there is a supportive rule $R$ that is not overruled,
each of whose body literals are proved defeasibly according to $Y$.

{\small
\begin{Clause}\label{defeasibly_free}
  $\mt{defeasibly_Z}(\free\ X)$ :-\\
  \> $\mt{proof\_tag}(Z)$,\\
  \> $\mt{defeasibly_Z}(Z\ X)$.
\end{Clause}

\begin{Clause}\label{defeasibly_fail}
  $\mt{defeasibly_Z}(\fail\ X)$ :-\\
  \> $\mt{not\ defeasibly_Z}(X)$.
\end{Clause}

\begin{Clause}\label{defeasibly1}
  $\mt{defeasibly_Z}(X)$ :-\\
  \> $\mt{definitely}(X)$.
\end{Clause}

\begin{Clause}\label{defeasibly_notfree}
  $\mt{defeasibly_Z}(Y\ X)$ :-\\
  \> $\mt{proof\_tag}(Y)$,\\
%  \> $ Y \neq \fail$, $ Y \neq \free$, \\
  \> $\mt{not\ definitely}(\non X)$,\\
  \> $\mt{supportive\_rule}(R,X,[\seq{W}])$,\\
  \> $\mt{defeasibly_Y}(W_1)$,\dots,$\mt{defeasibly_Y}(W_n)$,\\
  \> $\mt{not\ overruled_Y}(R,X)$.
\end{Clause}
}

The basic structure of overruling a rule is similar for all defeasible logics:
the body of the overruling rule must be proved and the rule not ``defeated''.
However,
it varies depending on whether the logic blocks or propagates ambiguity.
In an ambiguity blocking logic, the body of the overruling rule must be established defeasibly
whereas,
in an ambiguity propagating logic, the body of the overruling rule need only be supported.

{\small
\begin{Clause}\label{overruled_AB}
  $\mt{overruled_Z}(R,X)$ :-\\
  \> $\mt{ambiguity\_blocking}(Z)$, \\
  \> $\mt{rule}(S,\non X,[\seq{U}])$,\\
  \> $\mt{defeasibly_Z}(U_1)$,\dots,$\mt{defeasibly_Z}(U_n)$,\\
  \> $\mt{not\ defeated_Z}(R,S,\non X)$.
\end{Clause}

\begin{Clause}\label{overruled_AP}
  $\mt{overruled_Z}(R,X)$ :-\\
  \> $\mt{ambiguity\_propagating}(Z)$, \\
  \> $\mt{rule}(S,\non X,[\seq{U}])$,\\
  \> $\mt{supported_Z}(U_1)$,\dots,$\mt{supported_Z}(U_n)$,\\
  \> $\mt{not\ defeated_Z}(R,S,\non X)$.
\end{Clause}
}

The notion of defeat varies, depending on whether a logic involves team defeat or individual defeat.
In individual defeat, the overruling rule $S$ is defeated if the rule $R$ it tries to overrule is superior to $S$.
In team defeat, $S$ is defeated if there is a rule $T$ (possibly the same as $R$) that is superior to $S$ and
whose body can be proved.

{\small
\begin{Clause}\label{defeated_team}
  $\mt{defeated_Z}(R,S,\non X)$ :-\\
  \> $\mt{team}(Z)$, \\
  \> $\mt{sup}(T,S)$, \\
  \> $\mt{supportive\_rule}(T,X,[\seq{V}])$,\\
  \> $\mt{defeasibly_Z}(V_1)$,\dots,$\mt{defeasibly_Z}(V_n)$.
\end{Clause}

\begin{Clause}\label{defeated_indiv}
  $\mt{defeated_Z}(R,S,\non X)$ :-\\
  \> $\mt{indiv}(Z)$, \\
  \> $\mt{sup}(R,S)$. 
\end{Clause}
}

The structure of this meta-program makes one point clear that was less readily apparent in \cite{flexf} or \cite{TOCL10}:
treatment of ambiguity concerns how the body of an overruling rule is proved,
while the choice of team/individual defeat concerns how an overruling rule can be defeated.

For the ambiguity propagating logics we must define the notion of ``supported''.
The intuition is that a literal is supported if there is a chain of supportive rules that form a proof tree
for the literal, and each supportive rule is not beaten (i.e. overruled) by a rule that is proved defeasbily.
In ordinary defeasible logics support is only needed for the ambiguity propagating logics
but, for annotated defeasible theories, we also need to have support for ambiguity blocking logics.
This is because we might wish to use, as part of the support, 
a rule that contains an annotated literal such as $\partial q$.
Hence the $\mt{supported}$ predicate is defined uniformly, 
with a parameter $\mt{Z}$ specifying the form of defeasibility underlying the support.
Thus we are introducing new forms of support: $\supp_\partial$ and $\supp_{\partial^*}$.

As with $\mt{defeasibly}$, the clauses for $\mt{supported}$ address
free literals,
fail-expressions,
literals that are proved definitely,
and proof-annotated literals.
Note how the parameter $\mt{Z}$ to $\mt{supported}$ is used by $\mt{beaten}$ to select the
form of defeasibility for which the body of an overruling rule must be proved.

{\small
\begin{Clause}\label{support_free}
  $\mt{supported_Z}(\free\ X)$ :-\\
%  \> $\mt{proof\_tag}(Z)$,\\
  \> $\mt{supported_Z}(Z\ X)$.
\end{Clause}

\begin{Clause}\label{support_fail}
  $\mt{supported_Z}(\fail\ X)$ :-\\
  \> $\mt{not\ supported_Z}(X)$.
\end{Clause}

\begin{Clause}\label{support_definite}
  $\mt{supported_Z}(X)$ :-\\
  \> $\mt{definitely}(X)$.
\end{Clause}

\begin{Clause}\label{support_notfree}
  $\mt{supported_Z}(Y\ X)$ :-\\
  \> $\mt{proof\_tag}(Y)$,\\
%  \> $ Y \neq \fail$, $ Y \neq \free$, \\
  \> $\mt{supportive\_rule}(R, X, [\seq{W}])$,\\
  \> $\mt{supported_Y}(W_1)$,\dots,$\mt{supported_Y}(W_n)$,\\
  \> $\mt{not\ beaten_Y}(R,X)$.
\end{Clause}

\begin{Clause}\label{beaten1}
  $\mt{beaten_Z}(R,X)$ :-\\
  \>$\mt{rule}(S,\non X,[\seq{W}])$,\\
  \>$\mt{defeasibly_Z}(W_1)$,\dots,$\mt{defeasibly_Z}(W_n)$,\\
  \>$\mt{sup}(S,R)$.
\end{Clause}
}

Let us now examine how to put annotated defeasible logic to work by
revisiting the compensation example presented in the introduction. 
As we have already discussed, $\mi{Guilty}$ must be proven
with the ``beyond reasonable doubt'' proof standard to derive that 
the defendant is entitled to receive a compensation.

As we have alluded to in the introduction, \cite{GordonWalton:proof} 
proposed to model proof standards such as scintilla of evidence, 
preponderance of evidence, clear and convincing case, beyond reasonable 
doubts and  dialectical validity using rule based argumentation. For 
example, they define that the proof standard of preponderance of 
evident for a literal $p$ is satisfied if and only if the maximum weight 
of applicable arguments for $p$ exceeds some threshold $\alpha$, and the 
difference between the maximum weight of the applicable arguments for $p$
and the maximum weight of the applicable arguments against $p$ exceeds some 
threshold $\beta$. \cite{icail2011carneades} shows how the weights and 
thresholds can be modelled by a preference relation (superiority) over
arguments (rules) and it establishes the following relationships between the 
proof standards and proof tags:
\begin{center}
$\begin{array}{lc}
    \text{Proof standard(s)} & \text{Proof tag}\\
    \hline
    \text{scintilla of evidence}     & \sigma\\
    \text{preponderance of evidence, clear and convincing case} & \partial^{*}\\
    \text{beyond reasonable doubt, dialectic validity} & \delta^{*}
  \end{array}$
\end{center}
where the distinction between preponderance of evidence and clear and convincing case, and 
beyond reasonable doubt and dialectic validity depends on how the weights associated 
to the arguments and thresholds are translated in instances of the superiority relation 
in the resulting theories.  Furthermore, \cite{icail2011carneades} provides examples
where the definitions of proof standards given in \cite{GordonWalton:proof} exhibit
some counter-intuitive conclusions. To obviate such limitations he proposes an
alternative correspondence between proof tags in defeasible logic variants
and proof standards, including the following:
\begin{center}
$\begin{array}{lll}
    \text{Proof standard(s)} & ~~~ & \text{Proof tag}\\
    \hline
    \text{substantial evidence}     & & \sigma\\
    \text{preponderance of evidence} & & \partial\\
    \text{beyond reasonable doubt} & & \delta\\
    \text{dialectic validity} & & \delta \text{ (when the superiority relation is ignored)}
  \end{array}$
\end{center}

Thus, the proof standard of beyond reasonable doubt corresponds to
defeasible provability using ambiguity propagation.  
Accordingly, we can replace $\mi{BeyondReasonableDoubt}$ in rule $r'_5$ with $+\delta$.
All the other literals appearing in the body of the rules do not 
require special proof standards, and thus we can annotate them with 
$\free$. Consequently, the formalization of this scenario in 
annotated defeasible logic is: 
\[
\begin{array}{lrcl}
  r_1\colon & \free\:\mi{EvidenceA} & \Rightarrow &\neg\mi{Responsible}\\
  r_2\colon & \free\:\mi{EvidenceB} & \Rightarrow &\phantom{\neg}\mi{Responsible}\\
  r_3\colon & \free\:\mi{Responsible} & \Rightarrow  & \phantom{\neg}\mi{Guilty}\\
  r_4\colon && \Rightarrow & \neg\mi{Guilty} \\  
  r_5\colon & +\delta\neg\:{Guilty} &\Rightarrow &\phantom{\neg}\mi{Compensation}\\
  r_6\colon & & \Rightarrow & \neg\mi{Compensation}
\end{array}
\]
It is easy to verify that we now derive $+\partial \neg\mi{Compensation}$, that the defendant
is not entitled to compensation, as the scenario requires.  

\section{Properties of Annotated Defeasible Theories}

We now investigate properties of annotated defeasible logic,
exploiting its logic programming underpinnings.

The first theorem relates the meta-program for annotated defeasible logic
to the meta-programs for existing defeasible logics $\DL(d)$.
Those logics do not contain fail-expressions.
We write $\models_K$ for logical consequence under Kunen's semantics \cite{Kunen}.
Recall that $\cM_d(D)$ is the meta-programming representation for $D$ in $\DL(d)$,
while $\cM(\alpha(D))$ is the meta-programming representation for $D$ annotated by $\alpha$.

\begin{theorem}  \label{thm:correct}
Let $D = (F, R, >)$ be a defeasible theory, and $\alpha$ be an annotation function for that theory.
Let $d \in \{ \delta^*, \delta, \partial^*, \partial \}$. 
% \supp_\partial, \supp_{\partial^*}, \supp_\delta, \supp_{\delta^*} \}$.

Suppose $\alpha(R)$ contains only annotations $\free$ and $d$, and there is no fail-expression in $R$. 
Then, for every literal $q$

\begin{itemize}
\item
$\cM(\alpha(D)) \models_K \mt{defeasibly}_d(d~q)$ iff 
$\cM_d(D) \models_K \mt{defeasibly}(q)$

\item
$\cM(\alpha(D)) \models_K \neg \mt{defeasibly}_d(d~q)$ iff 
$\cM_d(D) \models_K \neg \mt{defeasibly}(q)$
\end{itemize}

\noindent
Furthermore, if $d \in \{ \delta^*, \delta \}$,
\begin{itemize}
\item
$\cM(\alpha(D)) \models_K \mt{supported}_d(d~q)$ iff 
$\cM_d(D) \models_K \mt{supported}(q)$
\item
$\cM(\alpha(D)) \models_K \neg \mt{supported}_d(d~q)$ iff 
$\cM_d(D) \models_K \neg \mt{supported}(q)$
\end{itemize}

\end{theorem}

The proof is based on separately unfolding $\cM(\alpha(D))$ and $\cM_d(D)$
until they have essentially the same form.  

As an immediate corollary to this theorem, we see that annotated defeasible theories are a conservative extension of defeasible theories.
Let the \emph{free annotation function} be the annotation function that maps every body literal occurrence in $D$ to $\free$.
For any defeasible theory $D$,
the unannotated theory behaves exactly the same as the theory annotated by the free annotation function.

\begin{corollary}  \label{cor:consext}
Suppose that $\alpha_F$ is the free annotation function for $D$.
Let $d \in \{ \delta^*, \delta, \partial^*, \partial \}$. 
Then,
for every literal $q$, 

\begin{itemize}
\item
$\cM(\alpha_F(D)) \models_K \mt{defeasibly_d}(q)$ iff $D \vdash +d q$ 
\item
$\cM(\alpha_F(D)) \models_K \neg \mt{defeasibly_d}(q)$ iff $D \vdash -d q$
\end{itemize}

\noindent
Furthermore, if $d \in \{ \delta^*, \delta \}$,
\begin{itemize}
\item
$\cM(\alpha_F(D)) \models_K \mt{supported_d}(q)$ iff $D \vdash +\supp_d q$ 
\item
$\cM(\alpha_F(D)) \models_K \neg \mt{supported_d}(q)$ iff $D \vdash -\supp_d q$
\end{itemize}

\end{corollary}

For any tag $d$ and an annotated defeasible theory $D$ we define
${+}d(D) = \{ q ~|~  D \vdash {+}d q\} = \{ q ~|~ \cM(D) \models_K \mt{defeasibly_d}(q) \}$ and
${-}d(D) = \{ q ~|~  D \vdash {-}d q\} = \{ q ~|~ \cM(D) \models_K \neg \mt{defeasibly_d}(q) \}$.
Similarly, we define
${+}\supp_d(D)$ as $\{ q ~|~ \cM(D) \models_K \mt{supported_d}(q) \}$ and
${-}\supp_d(D)$ as $\{ q ~|~ \cM(D) \models_K \neg \mt{supported_d}(q) \}$.

We can now extend the inclusion theorem of \cite{TOCL10} to the new tags and annotated defeasible logic.
This theorem shows the relative inference strength of the different forms of defeasibility.

\begin{theorem}[Inclusion Theorem]  \label{thm:inclusion} 
Let $D$ be an annotated defeasible theory.
\begin{itemize}
\item[(a)] ${+}\Delta(D) \subseteq {+}\delta^*(D) \subseteq {+}\delta(D) \subseteq {+}\partial(D)
\subseteq {+}\supp_\delta(D) \subseteq {+}\supp_{\delta^*}(D)$

\item[(b)] $-\supp_{\delta^*}(D) \subseteq -\supp_\delta(D) \subseteq -\partial(D) \subseteq -\delta(D)
\subseteq -\delta^*(D) \subseteq -\Delta(D)$

\item[(c)] ${+}\partial(D) \subseteq {+}\supp_\partial(D) \subseteq {+}\supp_\delta(D)$

\item[(d)] $-\supp_\delta(D) \subseteq {-}\supp_\partial(D) \subseteq {-}\partial(D)$

\item[(e)] ${+}\delta^*(D) \subseteq {+}\partial^*(D) \subseteq {+}\supp_{\partial^*}(D) \subseteq {+}\supp_{\delta^*}(D)$

\item[(f)] $-\supp_{\delta^*}(D) \subseteq {-}\supp_{\partial^*}(D) \subseteq {-}\partial^*(D) \subseteq -\delta^*(D)$
\end{itemize}
\end{theorem}

The proof is by induction on the iteration stages of Fitting's $\Phi_{\cM(D)}$ function.

The inclusions in this theorem are presented graphically in Figure \ref{fig:inclusion}.
The relation $t_1 \subset t_2$ expresses that,
for all defeasible theories $D$, $+t_1(D) \subseteq +t_2(D)$ and $-t_1(D) \supseteq -t_2(D)$,
and, for some defeasible theory $D$, $+t_1(D) \subset +t_2(D)$.
The containments come from the theorem, while their strictness is demonstrated by simple examples.
Examples also show that there are no containments that can be added to the figure.

\begin{figure*}
\[
\begin{array}{rcccl}
 \Delta   \subset       \delta^*          &  \subset  & \delta    \subset  \partial   \subset  \supp_\partial  \subset  \supp_\delta &  \subset & \supp_{\delta^*} \\
 \\
                                                             &   \sesubset &                                                                               & \nesubset & \\
 \\
                                                            &                     &      \partial^*  ~~~~ \subset ~~~~ \supp_{\partial^*}        &\\
\end{array}
\]

\caption{Ordering of inference rules by relative inference strength.}
\label{fig:inclusion}
\end{figure*}

This ordering on tags can be extended to annotation functions.
Let $\alpha_1$ and $\alpha_2$ be annotation functions for a defeasible theory $D$.
We define $\alpha_1 \sqsubseteq \alpha_2$ iff for every body occurrence $o$ of every literal in $D$, 
$\alpha_1(o) \subset \alpha_2(o)$.
If such an ordering had implications for the conclusions of the annotated theories, it would provide
a useful basis from which to reason about annotated defeasible theories.
Unfortunately, the most obvious possibility -- a kind of monotonicity --
does not hold, as the following example shows.
 
\begin{example}
Let $D$ consist of the rules
\[
\begin{array}{lrclclrcl}
r_1: &           & \Rightarrow & \phantom{\neg} p & ~~~~~~~~~~~~~ &
r_5: &     q    & \Rightarrow & \phantom{\neg} s \\
r_2: &           & \Rightarrow & \neg p & &
r_6: &           & \Rightarrow & \neg s \\
r_3: &           & \Rightarrow & \phantom{\neg} q \\
r_4: &  \neg p & \Rightarrow & \neg q \\
\end{array}
\]
with $r_5 > r_6$.

Let $\alpha_1$ map $q$ in $r_5$ to $\delta$, and $\alpha_2$ map $q$ in $r_5$ to $\partial$
(with all other occurrences mapped to $\free$).
Then $\alpha_1 \sqsubseteq \alpha_2$.
Rules $r_1$ - $r_4$ are a standard example distinguishing ambiguity blocking and propagating behaviours.
$+\partial q$ and $-\delta q$ can be concluded.
Consequently, in $\alpha_1(D)$ we conclude $+\partial \neg s$ and $-\partial s$
while in $\alpha_2(D)$ we conclude $-\partial \neg s$ and $+\partial s$.

Thus we see that a strengthening of the annotation function (in the $\sqsubseteq$ ordering)
does not necessarily lead to a strengthening of the conclusions
of the annotated defeasible theory.
\end{example}

For the defeasible logics we address, the consequences of a defeasible theory
can be computed in linear time, with respect to the size of the theory \cite{Maher2001,TOCL10},
but these logics only support one form of defeasibility.
Annotated defeasible logic allows the interaction between the different inference rules
but, nevertheless, we expect its consequences can also be computed in linear time,
although with a larger constant factor.
(Certainly, it is straightforward to show we can compute consequences in quadratic time.
See the supplementary material.)
% Just unfold everything, which gives us an essentially proposition theory of quadratic size in the worst case.
% The complexity of computing consequences in Kunen's semantics is then linear, giving O($|D|^2$).
Let 
\[
\begin{array}{rl}
\cC(D) = 
& \{ {+}d q ~|~ \cM(D) \models_K \mt{defeasibly}_d(q), d \in T \} ~\cup \\
& \{ {-}d q ~|~ \cM(D) \models_K \neg \mt{defeasibly}_d(q), d \in T \} ~\cup \\
& \{ {+}\supp_d q ~|~ \cM(D) \models_K \mt{supported}_d(q), d \in T \} ~\cup \\
& \{ {-}\supp_d q ~|~ \cM(D) \models_K \neg \mt{supported}_d(q), d \in T \}  \\
\end{array}
\]
where $D$ is an annotated defeasible theory,
$T = \{\partial, \partial^*, \delta, \delta^*\}$ refers to the four main forms of defeasibility,
and $q$ ranges over annotated literals.

\begin{conjecture}  \label{thm:linear}
Let $D$ be an annotated defeasible theory, and $|D|$ be the number of symbols in $D$.
Then the set of consequences $\cC(D)$ can be computed in time O($|D|$).
\end{conjecture}

\section{Different Forms of Failure}

One advantage of the framework of \cite{MG99,flexf} is that different notions of failure can be obtained
by different semantics for logic programs.
In this section we demonstrate that annotated defeasible logic is a conservative extension of those logics for many such semantics.

Many of the logic programming semantics we will focus on can be seen to be derived from the
3-valued stable models \cite{Pmodels} (also known as \emph{partial stable models},
but distinct from partial stable models in \cite{SaccaZaniolo}).
In addition to the semantics based on all partial stable models,
there is the well-founded model \cite{WF91},  
which is the least partial stable model under the information ordering \cite{Pmodels} (called $F$-least in \cite{Pmodels});
the (2-valued) stable models \cite{stable};
the regular models \cite{regular}, 
which are the maximal partial stable models under set inclusion on the positive literals;
and
the L-stable models \cite{Lstable},
which are the maximal partial stable models under set inclusion on positive and negative literals or,
equivalently,
the minimal partial stable models under set inclusion on the undefined literals.
The interest in these semantics derives from the use of their counterparts
in abstract argumentation \cite{LPeqArg}.

Let $\cS$ denote the collection of semantics mentioned above, with the exception of the stable semantics.
That is, $\cS = \{ \mi{partial~stable, well\mbox{--}founded, regular, L\mbox{--}stable, Kunen, Fitting} \}$.
These semantics (and the stable semantics) are preserved by unfolding
(see \cite{Aravindan,cdr}).
Consequently, Theorem \ref{thm:correct} extends to the semantics in $\cS$:

\begin{theorem}  \label{thm:correct2}
Let $D = (F, R, >)$ be a defeasible theory, and $\alpha$ be an annotation for that theory.
Let $d \in \{ \delta^*, \delta, \partial^*, \partial \}$. 
% \supp_\partial, \supp_{\partial^*}, \supp_\delta, \supp_{\delta^*} \}$.
Suppose $\alpha(R)$ contains only annotations $\free$ and $d$, and there is no fail-expression in $R$. 
Let $S \in \cS$.
Then

\begin{itemize}
\item
$\cM(\alpha(D)) \models_S \mt{defeasibly_d}(q)$ iff 
$\cM_d(D) \models_S \mt{defeasibly}(q)$

\item
$\cM(\alpha(D)) \models_S \neg \mt{defeasibly_d}(q)$ iff 
$\cM_d(D) \models_S \neg \mt{defeasibly}(q)$
\end{itemize}

\noindent
and, if $d \in \{ \delta^*, \delta \}$,
\begin{itemize}
\item
$\cM(\alpha(D)) \models_S \mt{supported_d}(q)$ iff 
$\cM_d(D) \models_S \mt{supported}(q)$
\item
$\cM(\alpha(D)) \models_S \neg \mt{supported_d}(q)$ iff 
$\cM_d(D) \models_S \neg \mt{supported}(q)$
\end{itemize}

More generally,
the S-models of $\cM(\alpha(D))$ restricted to $\mt{defeasibly_d}$ are identical (up to predicate renaming) 
to the S-models of $\cM_d(D)$ restricted to $\mt{defeasibly}$.
\end{theorem}

In particular, annotated defeasible logic under the well-founded semantics
extends the well-founded defeasible logics \cite{MG99,WFDL}.

This theorem does \emph{not} apply to the stable model semantics, because of the possibility that
$\cM_d(D)$ has stable models but $\cM(D)$ does not.
This, in turn, occurs because $\cM(D)$ represents all the inference rules, while $\cM_d(D)$ does not.
Technically, the proof fails because the deletion of irrelevant clauses is not sound under the stable model semantics.
To see what can go wrong, consider the following example.

\begin{example}
Let $D$ consist of the rules
\[
\begin{array}{lrclclrcl}
r_1: &           & \Rightarrow & \phantom{\neg} p & ~~~~~~~~~~~&
r_3: &           & \Rightarrow & \phantom{\neg} q \\

r_2: &  p, q   & \Rightarrow & \neg p & &
r_4: &           & \Rightarrow & \phantom{\neg} q \\
& & & & &
r_5: &           & \Rightarrow & \neg q \\
& & & & &
r_6: &           & \Rightarrow & \neg q \\
\end{array}
\]
with $r_3 > r_5$ and $r_4 > r_6$.

After unfoldings and simplifications,
$\cM(D)$ contains

{\small
\begin{Clause}
  $\mt{defeasibly_\partial}(\partial\ p)$ :-\\
  \> $\mt{not\ overruled_\partial}(r_1, p)$.
\end{Clause}
\begin{Clause}
  $\mt{overruled_\partial}(r_1, p)$ :-\\
  \> $\mt{defeasibly_\partial}(\partial\ p)$, \\
  \> $\mt{defeasibly_\partial}(\partial\ q)$.
\end{Clause}
}
\noindent
and similar clauses for $\partial^*$
(as well as other clauses).   % derived from $r_1$-$r_2$ and clauses derived from $r_3$-$r_6$.

It is clear that if $\mt{defeasibly_\partial}(\partial\ q)$ holds then the structure of these two clauses
prevents the existence of a stable model,
while if $\neg \mt{defeasibly_\partial}(\partial\ q)$ then $\mt{defeasibly_\partial}(\partial\ p)$ holds in every stable model,
assuming there is nothing else preventing the formation of stable models.
The same applies for $\partial^*$.

Now, $\mt{defeasibly_\partial}(\partial\ q)$ holds, but $\mt{defeasibly_{\partial^*}}(\partial^*\ q)$ does not.
It follows, from the proof of Theorem \ref{thm:correct},
that $\cM_{\partial^*}(D)$ has stable models but $\cM(D)$ does not.
\end{example}

Thus Theorem \ref{thm:correct2} holds for stable models only when \emph{all} forms of defeasibility
and supportedness have stable models.

\section{Related Work}

Among the features of annotated defeasible theories are:
(1)
the language supports multiple forms of defeasibility within a single defeasible theory, 
indeed within a single rule; 
(2)  the language provides explicit fail-expressions;
(3) the framework has the ability to incorporate different notions of failure-to-prove,
corresponding to different semantics of negation-as-failure.
No other formalism for defeasible reasoning has all these features.

Courteous logic programs \cite{Grosof97} (and later developments \cite{LPDA,ASPDA})
permit negation-as-failure expressions in defeasible rules, which are essentially the same as fail-expressions.
\cite{LPwN} discussed a specific transformation for eliminating these expressions from courteous logic programs;
that transformation is not sound for ambiguity propagating logics.
%(It is also known that the transformation for eliminating the superiority relation in $DL(\partial)$ \cite{TOCL01}
%is not correct for ambiguity propagating logics \cite{Lam.2011}.)
Our meta-programming approach to fail-expressions was discussed in \cite{MG99}, for a language with a single form of defeasibility, and our Theorem \ref{thm:correct} extends to languages with such fail-expressions.

Within proof-theoretic treatments of defeasible logics
(see, for example \cite{MN10} and \cite{TOCL10})
the logics can incorporate multiple forms of defeasibility, but they don't interact.
For example, the proof of $+\partial q$ cannot depend on the proof of $+\delta p$:
it can only depend on proofs of $\partial$ conclusions.
Within the meta-programming framework of \cite{MG99,flexf}
a logic has only a single form of defeasibility,
although this can be easily remedied by the use of multiple variants of the $\mt{defeasibly}$ predicate.
Still, the multiple forms don't interact.
Structured argumentation approaches, such as ASPIC+ \cite{ASPIC+},
use unannotated rules without an inference rule (in the sense above)
and hence define a single form of defeasibility.
A meta-program component of the languages LPDA and ASPDA \cite{LPDA,ASPDA},
called an argumentation theory,
is capable of specifying a different inference rule for each literal,
but not for each \emph{occurrence} of each literal.
Thus, although they provide more interaction than the defeasible logics,
they do not provide the ability to apply different inference rules to the same atom.

It should be noted that the logics of \cite{TOCL10}
are able to simulate each other \cite{Maher12,Maher13}
(and ASPIC+ appears expressive enough to simulate these logics),
but such an approach to incorporating multiple forms of defeasibility leads to an unnatural representation
and has computational penalties.
It also fails to represent free-expressions,
since the top level form of defeasibility must be fixed before simulations can be coded.

Annotated logic programs \cite{GALP}
are an extension of logic programs to multi-valued logics,
where the truth values are assumed to form an upper semi-lattice.
Atoms in the body are annotated by truth values
and the head is annotated by a function of those truth values.
Thus there are some similarities to annotated defeasible logic,
in the use of annotations, including a similarity of 
variable annotations and free-expressions.
However, annotated defeasible logic uses proof tags -- not truth values -- as annotations,
and does not assume any ordering on the annotations.
Further, the semantics of annotated logic programs is essentially a disjunction of the conclusions of rules,
so this formalism is unable to represent the overriding of a rule by a competing rule.

Most defeasible logics support a single semantics of failure:
Kunen's \cite{TOCL10}, well-founded \cite{MG99,MN10,WFDL,Grosof97,LPDA}, stable \cite{DefLog,Maier13,ASPDA}.
Apart from the framework of \cite{flexf}, the only defeasible formalisms supporting multiple semantics
are structured argumentation languages like ASPIC+ \cite{ASPIC+}.
But such languages do not support multiple forms of defeasibility.

The annotation mechanism  we presented is closely related to the introduction
of modal literals in modal defeasible logic \cite{tplp:goal}, where each rule is labelled with the mode ($\Box$)
its conclusion can be proved and the literals $\Box q$ and $\Diamond q$ correspond to 
$+\partial_{\Box} q$ and $-\partial_{\Box}\neg q$. While each modality has its own inference
rule, each supports a single form of defeasibility. This raised the question whether different
forms of defeasibility could be combined: the present paper offers a positive answer.

\section{Conclusion}

We have argued that we need a formalism that supports different kinds of defeasible reasoning,
and introduced annotated defeasible logic to fulfil that requirement.
%We demonstrated the use of annotated defeasible logic to integrate the logics addressed in \cite{TOCL10},
%but the same approach can be used to integrate other defeasible logics.
The semantics of the annotated logic is defined through a logic program,
and we are able to exploit that medium to prove properties of the logic.

%\newpage
\bibliographystyle{acmtrans}
\bibliography{ann_DL}

\begin{appendix}

\section{Inference Rules}

Defeasible logics are usually defined via their proof mechanism.
Here we present the inference rules for the four defeasible logics we integrate within
annotated defeasible logic.
Each inference rule is labelled by the kind of conclusions it infers.
The presentation is adapted from \cite{TOCL10}.
A defeasible logic is determined by the inference rules it allows.
For example, 
$\DL(\partial)$ allows $+\partial$ and ${-}\partial$, while
$\DL(\delta)$ allows $+\delta$, ${-}\delta$, $+\sigma_\delta$, and ${-}\sigma_\delta$.

A proof $P$ is a sequence of conclusions.
The conclusion at position $i$ in the sequence is denoted by $P(i)$,
and a prefix of the proof of length $i$ is denoted by $P[1..i]$.
The inference rules establish when a conclusion can be drawn at position $i+1$,
given the conclusions already proved ($P[1..i]$).
Where $q$ is a literal,
$R_{sd}[q]$ denotes the set of strict or defeasible rules with head $q$, while
$R[q]$ denotes the set of all rules (including defeaters) with head $q$.
For a rule $r$, $A(r)$ denotes the antecedent (or body) of $r$.

\smallskip
\noindent\begin{minipage}[t]{.45\textwidth}
\begin{tabbing}
$+\partial)$  Infer  $P(i+1) = +\partial q$  if either \\
\hspace{0.2in}  .1)  $+\Delta q \in P[1..i]$; or  \\
\hspace{0.2in}  .2)  The following three conditions all hold. \\
\hspace{0.4in}      .1)  $\exists r \in R_{sd}[q] \  \forall a \in A(r),  +\partial a \in P[1..i]$,  and \\
\hspace{0.4in}      .2)  $-\Delta \non q \in P[1..i]$,  and \\
\hspace{0.4in}      .3)  $\forall s \in R[\non q]$  either \\
\hspace{0.6in}          .1)  $\exists a \in A(s),  -\partial a \in P[1..i]$;  or \\
\hspace{0.6in}          .2)  $\exists t \in R_{sd}[q]$  such that \\
\hspace{0.8in}              .1)  $\forall a \in A(t),  +\partial a \in P[1..i]$,  and \\
\hspace{0.8in}              .2)  $t > s$.
\end{tabbing}
\end{minipage}
\begin{minipage}[t]{.45\textwidth}
\begin{tabbing}
$-\partial)$  Infer  $P(i+1) = -\partial q$  if \\
\hspace{0.2in}  .1)  $-\Delta q \in P[1..i]$, and \\
\hspace{0.2in}  .2)  either \\
\hspace{0.4in}      .1)  $\forall r \in R_{sd}[q] \  \exists a \in A(r),  -\partial a \in P[1..i]$; or \\
\hspace{0.4in}      .2)  $+\Delta \non q \in P[1..i]$; or \\
\hspace{0.4in}      .3)  $\exists s \in R[\non q]$  such that \\
\hspace{0.6in}          .1)  $\forall a \in A(s),  +\partial a \in P[1..i]$,  and \\
\hspace{0.6in}          .2)  $\forall t \in R_{sd}[q]$  either \\
\hspace{0.8in}              .1)  $\exists a \in A(t),  -\partial a \in P[1..i]$;  or \\
\hspace{0.8in}              .2)  not$(t > s)$.\\
\end{tabbing}
\end{minipage}

\smallskip
\noindent\begin{minipage}[t]{.45\textwidth}
\begin{tabbing}
$+\delta)$  Infer  $P(i+1) = +\delta q$  if either \\
\hspace{0.2in}  .1)  $+\Delta q \in P[1..i]$; or  \\
\hspace{0.2in}  .2)  The following three conditions all hold. \\
\hspace{0.4in}      .1)  $\exists r \in R_{sd}[q] \  \forall a \in A(r),  +\delta a \in P[1..i]$,  and \\
\hspace{0.4in}      .2)  $-\Delta \non q \in P[1..i]$,  and \\
\hspace{0.4in}      .3)  $\forall s \in R[\non q]$  either \\
\hspace{0.6in}          .1)  $\exists a \in A(s),  -\sigma_\delta a \in P[1..i]$;  or \\
\hspace{0.6in}          .2)  $\exists t \in R_{sd}[q]$  such that \\
\hspace{0.8in}              .1)  $\forall a \in A(t),  +\delta a \in P[1..i]$,  and \\
\hspace{0.8in}              .2)  $t > s$.
\end{tabbing}
\end{minipage}
\begin{minipage}[t]{.45\textwidth}
\begin{tabbing}
$-\delta)$  Infer  $P(i+1) = -\delta q$  if \\
\hspace{0.2in}  .1)  $-\Delta q \in P[1..i]$, and \\
\hspace{0.2in}  .2)  either \\
\hspace{0.4in}      .1)  $\forall r \in R_{sd}[q] \  \exists a \in A(r),  -\delta a \in P[1..i]$; or \\
\hspace{0.4in}      .2)  $+\Delta \non q \in P[1..i]$; or \\
\hspace{0.4in}      .3)  $\exists s \in R[\non q]$  such that \\
\hspace{0.6in}          .1)  $\forall a \in A(s),  +\sigma_\delta a \in P[1..i]$,  and \\
\hspace{0.6in}          .2)  $\forall t \in R_{sd}[q]$  either \\
\hspace{0.8in}              .1)  $\exists a \in A(t),  -\delta a \in P[1..i]$;  or \\
\hspace{0.8in}              .2)  not$(t > s)$.\\
\end{tabbing}
\end{minipage}

\begin{minipage}[t]{.45\textwidth}
\begin{tabbing}
$+\sigma_\delta)$  Infer  $P(i+1) = +\sigma_\delta q$  if either  \\
\hspace{0.2in}  .1)  $+\Delta q \in P[1..i]$; or\\
\hspace{0.2in}  .2)  $\exists r \in R_{sd}[q]$  such that  \\
\hspace{0.4in}      .1)  $\forall a \in A(r),  +\sigma_\delta a \in P[1..i]$,  and \\
\hspace{0.4in}      .2)  $\forall s \in R[\non q]$  either \\
\hspace{0.6in}          .1)  $\exists a \in A(s),  -\delta a \in P[1..i]$;  or \\
\hspace{0.6in}          .2)  not$(s > r)$.
\end{tabbing}
\end{minipage}
\begin{minipage}[t]{.45\textwidth}
\begin{tabbing}
$-\sigma_\delta)$  Infer  $P(i+1) = -\sigma_\delta q$  if \\
\hspace{0.2in}  .1)  $-\Delta q \in P[1..i]$, and \\
\hspace{0.2in}  .2)  $\forall r \in R_{sd}[q]$  either \\
\hspace{0.4in}      .1)  $\exists a \in A(r),  -\sigma_\delta a \in P[1..i]$;  or \\
\hspace{0.4in}      .2)  $\exists s \in R[\non q]$  such that \\
\hspace{0.6in}          .1)  $\forall a \in A(s),  +\delta a \in P[1..i]$,  and \\
\hspace{0.6in}          .2)  $s > r$.\\
\end{tabbing}
\end{minipage}

\begin{minipage}[t]{.45\textwidth}
\begin{tabbing}
$+\partial^{*})$  Infer  $P(i+1) = +\partial^{*}q$  if either  \\
\hspace{0.2in}  .1)  $+\Delta q \in P[1..i]$; or  \\
\hspace{0.2in}  .2)  $\exists r \in R_{sd}[q]$  such that \\
\hspace{0.4in}      .1)  $\forall a \in A(r),  +\partial^{*} a  \in  P[1..i]$,  and\\
\hspace{0.4in}      .2)  $-\Delta  \non q  \in  P[1..i]$,  and \\
\hspace{0.4in}      .3)  $\forall s  \in R[ \non q]$  either \\
\hspace{0.6in}          .1)  $\exists a \in A(s),  -\partial^{*}a  \in  P[1..i]$;  or \\
\hspace{0.6in}          .2)  $r > s$.
\end{tabbing}
\end{minipage}
\begin{minipage}[t]{.45\textwidth}
\begin{tabbing}
$-\partial^{*})$  Infer  $P(i+1) = -\partial^{*}q$  if \\
\hspace{0.2in}  .1)  $-\Delta q  \in  P[1..i]$, and \\
\hspace{0.2in}  .2)  $\forall r \in R_{sd}[q]$  either \\
\hspace{0.4in}      .1)  $\exists a \in A(r),  -\partial^{*}a  \in  P[1..i]$; or \\
\hspace{0.4in}      .2)  $+\Delta  \non q  \in  P[1..i]$; or \\
\hspace{0.4in}      .3)  $\exists s  \in R[ \non q]$  such that \\
\hspace{0.6in}          .1)  $\forall a \in A(s),  +\partial^{*}a  \in  P[1..i]$,  and \\
\hspace{0.6in}          .2)  not$(r > s)$. \\
\end{tabbing}
\end{minipage}

\begin{minipage}[t]{.45\textwidth}
\begin{tabbing}
$+\delta^{*})$  Infer  $P(i+1) = +\delta^{*}q$  if either \\
\hspace{0.2in}  .1)  $+\Delta q  \in  P[1..i]$; or  \\
\hspace{0.2in}  .2)  $\exists r  \in R_{sd}[q]$  such that \\
\hspace{0.4in}      .1)  $\forall a \in A(r),  +\delta^{*}a  \in  P[1..i]$,  and \\
\hspace{0.4in}      .2)  $-\Delta  \non q  \in  P[1..i]$,  and \\
\hspace{0.4in}      .3)  $\forall s  \in R[ \non q]$  either \\
\hspace{0.6in}          .1)  $\exists a \in A(s),  -\sigma_{\delta^*} ^{*}a  \in  P[1..i]$;  or \\
\hspace{0.6in}          .2)  $r > s$.
\end{tabbing}
\end{minipage}
\begin{minipage}[t]{.45\textwidth}
\begin{tabbing}
$-\delta^{*})$  Infer  $P(i+1) = -\delta^{*}q$  if \\
\hspace{0.2in}  .1)  $-\Delta q  \in  P[1..i]$, and \\
\hspace{0.2in}  .2)  $\forall r \in R_{sd}[q]$  either \\
\hspace{0.4in}      .1)  $\exists a \in A(r),  -\delta^{*}a  \in  P[1..i]$; or \\
\hspace{0.4in}      .2)  $+\Delta  \non q  \in  P[1..i]$; or \\
\hspace{0.4in}      .3)  $\exists s  \in R[ \non q]$  such that \\
\hspace{0.6in}          .1)  $\forall a \in A(s),  +\sigma_{\delta^*} ^{*}a  \in  P[1..i]$,  and \\
\hspace{0.6in}          .2)  not$(r > s)$.\\
\end{tabbing}
\end{minipage}

\begin{minipage}{.45\textwidth}
\begin{tabbing}
$+\sigma_{\delta^*} ^{*})$  Infer  $P(i+1) = +\sigma_{\delta^*} ^{*}q$  if either  \\
\hspace{0.2in}  .1)  $+\Delta q  \in  P[1..i]$; or  \\
\hspace{0.2in}  .2)  $\exists r  \in R_{sd}[q]$  such that  \\
\hspace{0.4in}      .1)  $\forall a \in A(r),  +\sigma_{\delta^*} ^{*}a  \in  P[1..i]$,  and \\
\hspace{0.4in}      .2)  $\forall s  \in R[ \non q]$  either \\
\hspace{0.6in}          .1)  $\exists a \in A(s),  -\delta^{*}a  \in  P[1..i]$;  or \\
\hspace{0.6in}          .2)  not$(s > r)$.
\end{tabbing}
\end{minipage}
\begin{minipage}{.45\textwidth}
\begin{tabbing}
$-\sigma_{\delta^*} ^{*})$  Infer  $P(i+1) = -\sigma_{\delta^*} ^{*}q$  if \\
\hspace{0.2in}  .1)  $-\Delta q  \in  P[1..i]$, and \\
\hspace{0.2in}  .2)  $\forall r \in R_{sd}[q]$  either \\
\hspace{0.4in}      .1)  $\exists a \in A(r),  -\sigma_{\delta^*} ^{*}a  \in  P[1..i]$;  or \\
\hspace{0.4in}      .2)  $\exists s  \in R[ \non q]$  such that \\
\hspace{0.6in}          .1)  $\forall a \in A(s),  +\delta^{*}a  \in  P[1..i]$,  and \\
\hspace{0.6in}          .2)  $s > r$.
\end{tabbing}
\end{minipage}

\section{Original Meta-programs}

The original metaprograms \cite{MG99,flexf} for the four main forms of defeasibility are outlined below.
They consist of clauses \ref{strictly1} and \ref{strictly2}, defining $\mt{definitely}$,
clauses
defining $\mt{rule}$ and $\mt{supportive\_rule}$
(see body of the paper),
and a selection of the following clauses for each form of defeasibility.

\setcounter{clause}{20}

{\small
\begin{Clause}\label{defeasibly1?}
  $\mt{defeasibly}(X)$ :-\\
  \> $\mt{definitely}(X)$.
\end{Clause}

\begin{Clause}\label{defeasibly2}
  $\mt{defeasibly}(X)$ :-\\
  \> $\mt{not\ definitely}(\non X)$,\\
  \> $\mt{supportive\_rule}(R,X,[\seq{Y}])$,\\
  \> $\mt{defeasibly}(Y_1)$,\dots,$\mt{defeasibly}(Y_n)$,\\
  \> $\mt{not\ overruled}(R,X)$.
\end{Clause}

\begin{Clause}\label{overruled}
  $\mt{overruled}(R,X)$ :-\\
%  \> $\mt{sup}(S,R)$,\\
  \> $\mt{rule}(S,\non X,[\seq{U}])$,\\
  \> $\mt{defeasibly}(U_1)$,\dots,$\mt{defeasibly}(U_n)$,\\
  \> $\mt{not\ defeated}(S,\non X)$.
\end{Clause}

\begin{Clause}\label{defeated}
  $\mt{defeated}(S,\non X)$ :-\\
  \> $\mt{sup}(T,S)$, \\
  \> $\mt{supportive\_rule}(T,X,[\seq{V}])$,\\
  \> $\mt{defeasibly}(V_1)$,\dots,$\mt{defeasibly}(V_n)$.
\end{Clause}
}

{\small
\begin{Clause}\label{support_trans1}
  $\mt{supported}(X)$ :-\\
  \> $\mt{definitely}(X)$.
\end{Clause}

\begin{Clause}\label{support_trans2}
  $\mt{supported}(X)$ :-\\
  \> $\mt{supportive\_rule}(R, X, [\seq{Y}])$,\\
  \> $\mt{supported}(Y_1)$,\dots,$\mt{supported}(Y_n)$,\\
  \> $\mt{not\ beaten}(R,X)$.
\end{Clause}

\begin{Clause}\label{beaten2}
  $\mt{beaten}(R,X)$ :-\\
  \>$\mt{rule}(S,\non X,[\seq{W}])$,\\
  \>$\mt{defeasibly}(W_1)$,\dots,$\mt{defeasibly}(W_n)$,\\
  \>$\mt{sup}(S,R)$.
\end{Clause}
}
{\small
\begin{Clause}\label{overruled2}
  $\mt{overruled}(R,X)$ :-\\
  \> $\mt{rule}(S,\non X,[\seq{U}])$,\\
  \> $\mt{supported}(U_1)$,\dots,$\mt{supported}(U_n)$,\\
  \> $\mt{not\ defeated}(S,\non X)$.
\end{Clause}
}
{\small
\begin{Clause}\label{noteams}
  $\mt{overruled}(R,X)$ :-\\
  \> $\mt{rule}(S,\non X,[\seq{U}])$,\\
  \> $\mt{defeasibly}(U_1)$,\dots,$\mt{defeasibly}(U_n)$,\\
  \> $\mt{not\ sup}(R,S)$.
\end{Clause}
}
{\small
\begin{Clause}\label{noteams2}
  $\mt{overruled}(R,X)$ :-\\
  \> $\mt{rule}(S,\non X,[\seq{U}])$,\\
  \> $\mt{supported}(U_1)$,\dots,$\mt{supported}(U_n)$,\\
  \> $\mt{not\ sup}(R,S)$.
\end{Clause}
}

The selection of clauses for each meta-program is as follows: \\
$\cM_\partial$ contains the clauses \ref{defeasibly1?} - \ref{defeated}. \\
$\cM_\delta$ contains the clauses \ref{defeasibly1?} - \ref{defeasibly2},  \ref{overruled2}, \ref{defeated},  and \ref{support_trans1} - \ref{beaten2}. \\
$\cM_{\partial^*}$ contains the clauses \ref{defeasibly1?} - \ref{defeasibly2}, and \ref{noteams}. \\
$\cM_{\delta^*}$ consists of the clauses  \ref{defeasibly1?} - \ref{defeasibly2}, \ref{noteams2},  and \ref{support_trans1} - \ref{beaten2}.

\section{Proofs of results}

We present (sketches of) proofs for the results in the paper.

%\newcounter{fig:inclusion}
%\setcounter{fig:inclusion}{1}

\setcounter{theorem}{1}
\addtocounter{theorem}{-1}

\begin{theorem}  \label{thm:correct}
Let $D = (F, R, >)$ be a defeasible theory, and $\alpha$ be an annotation function for that theory.
Let $d \in \{ \delta^*, \delta, \partial^*, \partial \}$. 
% \supp_\partial, \supp_{\partial^*}, \supp_\delta, \supp_{\delta^*} \}$.

Suppose $\alpha(R)$ contains only annotations $\free$ and $d$, and there is no fail-expression in $R$. 
Then, for every literal $q$

\begin{itemize}
\item
$\cM(\alpha(D)) \models_K \mt{defeasibly}_d(d~q)$ iff 
$\cM_d(D) \models_K \mt{defeasibly}(q)$

\item
$\cM(\alpha(D)) \models_K \neg \mt{defeasibly}_d(d~q)$ iff 
$\cM_d(D) \models_K \neg \mt{defeasibly}(q)$
\end{itemize}

\noindent
Furthermore, if $d \in \{ \delta^*, \delta \}$,
\begin{itemize}
\item
$\cM(\alpha(D)) \models_K \mt{supported}_d(d~q)$ iff 
$\cM_d(D) \models_K \mt{supported}(q)$
\item
$\cM(\alpha(D)) \models_K \neg \mt{supported}_d(d~q)$ iff 
$\cM_d(D) \models_K \neg \mt{supported}(q)$
\end{itemize}

\end{theorem}
\skipit{
\begin{proof}
(Sketch)
The proof of this theorem is similar for each tag $d$.
For brevity, we only provide the details for $\delta$.
The proof is based on unfolding $\cM(\alpha(D))$ until it has essentially the same form as an unfolding of $\cM_d(D)$.
The form of unfolding we use uses clauses from the current program,
and may be applied as long as no clause is used to unfold an atom in its own body.
%(no unfolding on direct recursion).
Such unfolding preserves the Kunen semantics (i.e. 3-valued models of the Clark-completion) of a logic program
by essentially the same argument that it preserves the 2-valued models \cite{trans}.
Clauses \ref{strictly1}, \ref{strictly2}, and \ref{defeasibly1} are the same in both $\cM$ and $\cM_d$,
so we will essentially ignore them.

In both $\cM(\alpha(D))$ and $\cM_\delta(D)$ 
we unfold all occurrences of the predicates used to represent the annotated defeasible theory,
and $\mt{rule}$ and $\mt{supportive\_rule}$.
Then, in $\cM(\alpha(D))$,
we unfold all occurrences of the predicates specifying the type of each tag:
$\mt{ambiguity\_propagating}$, $\mt{ambiguity\_blocking}$, $\mt{proof\_tag}$, $\mt{team}$, and $\mt{indiv}$.
At this point clauses derived from \ref{defeasibly_notfree}-\ref{overruled_AP} are ground, 
while clauses derived from \ref{defeated_indiv} only have a single, unused variable $X$ in their heads.
Similarly, clauses derived from \ref{support_notfree} and \ref{beaten1} are ground.

Then unfold all $\mt{defeasibly_Z}(\free ~ L)$ atoms.
This will not result in a clause unfolding itself: 
in \ref{defeasibly_free} because  $Z$ is not $\free$,
and
in  \ref{defeasibly_notfree} because  $Y$ is not $\free$.
Similarly, we unfold all $\mt{supported_Z}(\free ~ L)$ atoms.
As a result, $\free$ only occurs in the head of clauses derived from \ref{defeasibly_free} and \ref{support_free}.

Finally, unfold all  $\cM(\alpha(D))$ and $\mt{defeated}$ atoms in $\cM_\delta(D)$.
At this stage, clauses derived from $\cM_\delta(D)$ are essentially the same as 
some of the clauses derived from $\cM(\alpha(D))$ with subscript $\delta$;
the differences are in the name/arity of predicates (e.g., $\mt{defeasibly}$ versus $\mt{defeasibly_\delta}$)
and the presence of rules with heads of the form $\mt{defeasibly_\delta}(\free ~ L)$, $\mt{supported_\delta}(\free ~ L)$, $\mt{defeasibly_\delta}(\fail ~ L)$ or $\mt{supported_\delta}(\fail ~ L)$.
%The main difference is that $\mt{defeated_\delta}$ in $\cM(\alpha(D))$ has an extra argument $R$
%than $\mt{defeated_\delta}$ in $\cM_\delta(D)$, but $R$ is not used.
However, no atom with subscript $\delta$ depends on a predicate with a different subscript,
nor on clauses with $\free$ or $\fail$ in the head.
% Hence, we can apply a deletion transformation, viewing the clauses as essentially propositional,
% and deleting all clauses for atoms with a different subscript.
Hence, the consequences of $\cM(D)$ of the form $\mt{defeasibly_\delta}(q)$ and $\mt{supported_\delta}(q)$
are unaffected by the presence or absence of such rules, and so we delete them all.

Consequently, the two transformed programs are the same (modulo predicate renaming),
and hence have the same conclusions.
Since the transformations preserve the semantics of the programs, the result follows.
\end{proof}
}

\begin{corollary}  \label{cor:consext}
Suppose that $\alpha_F$ is the free annotation function for $D$.
Let $d \in \{ \delta^*, \delta, \partial^*, \partial \}$. 
Then,
for every literal $q$, 
% $d \in \{ \delta^*, \delta, \partial^*, \partial \}$
\begin{itemize}
\item
$\cM(\alpha_F(D)) \models_K \mt{defeasibly_d}(q)$ iff $D \vdash +d q$ 
\item
$\cM(\alpha_F(D)) \models_K \neg \mt{defeasibly_d}(q)$ iff $D \vdash -d q$
\end{itemize}

\noindent
Furthermore, if $d \in \{ \delta^*, \delta \}$,
\begin{itemize}
\item
$\cM(\alpha_F(D)) \models_K \mt{supported_d}(q)$ iff $D \vdash +\supp_d q$ 
\item
$\cM(\alpha_F(D)) \models_K \neg \mt{supported_d}(q)$ iff $D \vdash -\supp_d q$
\end{itemize}

\end{corollary}
\skipit{
\begin{proof}
The corollary follows from applying the previous theorem for each tag $d$ to the case where $\alpha$ is the free annotation function,
and the correctness of the individual meta-programs.
% but we don't have official proofs of correctness anywhere, apart from $\partial$
\end{proof}
}

\begin{theorem}[Inclusion Theorem]   \label{thm:inclusion} 
Let $D$ be an annotated defeasible theory.
\begin{itemize}
\item[(a)] ${+}\Delta(D) \subseteq {+}\delta^*(D) \subseteq {+}\delta(D) \subseteq {+}\partial(D)
\subseteq {+}\supp_\delta(D) \subseteq {+}\supp_{\delta^*}(D)$

\item[(b)] $-\supp_{\delta^*}(D) \subseteq -\supp_\delta(D) \subseteq -\partial(D) \subseteq -\delta(D)
\subseteq -\delta^*(D) \subseteq -\Delta(D)$

\item[(c)] ${+}\partial(D) \subseteq {+}\supp_\partial(D) \subseteq {+}\supp_\delta(D)$

\item[(d)] $-\supp_\delta(D) \subseteq {-}\supp_\partial(D) \subseteq {-}\partial(D)$

\item[(e)] ${+}\delta^*(D) \subseteq {+}\partial^*(D) \subseteq {+}\supp_{\partial^*}(D) \subseteq {+}\supp_{\delta^*}(D)$

\item[(f)] $-\supp_{\delta^*}(D) \subseteq {-}\supp_{\partial^*}(D) \subseteq {-}\partial^*(D) \subseteq -\delta^*(D)$
\end{itemize}
\end{theorem}
\skipit{
\begin{proof}
(Sketch)
Let $\Phi = \Phi_{\cM(D)}$ be Fitting's semantic function for the logic program $\cM(D)$ \cite{Fitting}.
Recall that Kunen's semantics is the set of all consequences of $\Phi \uparrow n$ for any finite $n$.
%Because a defeasible theory is finite, Fitting's semantics (the least fixedpoint of $\Phi$)
%is equal to Kunen's semantics.
We prove the containments by induction on the iteration of $\Phi$.
For brevity, we omit parts of the induction hypothesis related to proving (a).
We also omit the parts related to (b), (d) and (f) since, by the Principle of Strong Negation \cite{flexf},
their statements and proof are symmetric to those for the positive conclusions.
The induction hypothesis contains

\[
\begin{array}{rcll}
\mt{supported_{\partial^*}} \subseteq \mt{supported_{\delta^*}} & \wedge &
\mt{beaten_{\delta^*}} \subseteq \mt{beaten_{\partial^*}} & \wedge \\
\mt{supported_{\partial}} \subseteq \mt{supported_{\delta}} & \wedge &
\mt{beaten_{\delta}} \subseteq \mt{beaten_{\partial}} & \wedge \\
\mt{defeasibly_{\partial^*}} \subseteq \mt{supported_{\partial^*}} & \wedge &
\mt{beaten_{\partial^*}} \subseteq \mt{overruled_{\partial^*}}  &\wedge \\
\mt{defeasibly_{\delta^*}} \subseteq \mt{defeasibly_{\partial^*}} & \wedge &
\mt{overruled_{\partial^*}} \subseteq \mt{overruled_{\delta^*}} &\wedge \\
\mt{defeasibly_{\partial^*}} \subseteq \mt{supported_{\delta^*}} & \wedge &
\mt{beaten_{\delta^*}} \subseteq \mt{overruled_{\partial^*}}  & \\
%
%\mt{defeasibly_{\partial}} \subseteq \mt{supported_{\partial}} & \wedge &
%\mt{beaten_{\partial}|_q} \subseteq \mt{overruled_{\partial}|_q}  \\
%
\end{array}
\]

Clearly this statement holds in the empty interpretation.
It is mostly straightforward to show that if the induction hypothesis holds in $\Phi \uparrow n$
then it holds in $\Phi \uparrow {n{+}1}$.
For example, consider the first two containments in the induction hypothesis.
If they hold in $\Phi \uparrow n$
(and also $\mt{defeasibly_{\delta^*}} \subseteq \mt{defeasibly_{\partial^*}}$ holds)
then, applying clause \ref{beaten1}, the second containment holds in $\Phi \uparrow {n{+}1}$
and, applying clause \ref{support_notfree}, the first containment holds in $\Phi \uparrow {n{+}1}$.
To address fail-expressions we also need the corresponding versions of these containments and arguments
for negative conclusions.

One containment,  $\mt{defeasibly_{\partial}} \subseteq \mt{supported_{\partial}}$
is not easily proved by induction, but it has a direct proof.
For a set $S=\Phi \uparrow n$, if $\mt{defeasibly_{\partial}}(x) \in \Phi(S)$ then
there is a supportive rule $r$ whose body literals are defeasibly true in $S$
(i.e. $\mt{defeasibly_{\partial}}(w_i) \in S$)
and $\mt{not\ overruled_{\partial}}(r, x) \in S$.
Now $\mt{not\ overruled_{\partial}}(r, x) \in S$ only if,
for every rule $s$ for $\non x$ whose body literals are defeasibly true in $S$, 
there is a supportive rule $t$ whose body literals are defeasibly true in $S$ and $t > s$.
Since $D$ is finite and $>$ is acyclic,
for some such $t$,  for every such $s$, $s \not> t$.
This $t$ can now be used as $r$ in clauses \ref{support_notfree} and \ref{beaten1}
to show that $\mt{supported_{\partial}}(x) \in \Phi(S)$.
\end{proof}
}

As mentioned in the body of the paper, it is straightforward to compute the consequences
of an annotated defeasible theory in quadratic time.  We outline the proof.

\begin{propn}  \label{propn:quadratic}
Let $D$ be an annotated defeasible theory, and $|D|$ be the number of symbols in $D$.
Then the set of consequences $\cC(D)$ can be computed in time O($|D|^2$).
\end{propn}
\skipit{
\begin{proof}
(Sketch)
Consider the grounding of the clauses, by unfolding with the input representation of the defeasible theory
and related facts,
and the worst-case (i.e. maximum) size of the result.
Unfolding with facts like $\mt{proof\_tag}$ produces an increase in rules by a constant factor,
because the number of tags is fixed.
Unfolding clauses for $\mt{rule}$ etc.{} produces a set of ground instances linear in the size of rules in $D$.
For clauses \ref{strictly1} and \ref{strictly2}, the size of ground instances
is proportional to the size of facts/strict rules in $D$.
The size of ground instances of clauses \ref{defeasibly_notfree}, \ref{support_notfree}, and \ref{beaten1}
is proportional to the size of rules in $D$.
The size of ground instances of clauses \ref{defeated_indiv}
is proportional to the number of superiority statements in $D$.
The size of ground instances of clauses \ref{defeasibly_free} -- \ref{defeasibly1} and \ref{support_free} -- \ref{support_definite} is proportional to the number of literals in $D$.

For clauses \ref{overruled_AB} and \ref{overruled_AP},
the size of the ground instances is proportional to the product of the number of rules in $D$ and 
the maximum size of rules in $D$.
The size of ground instances of clauses \ref{defeated_team} is proportional to
 the product of the number of superiority statements and the maximum size of rules in $D$.

Thus the size of all ground clauses is bounded above by $|D|^2$.
The ground rules form an essentially propositional logic program.
Computing the consequences of a propositional logic program under the Kunen semantics
is linear in the size of the program.   % seems to be folklore
Consequently, the cost of computing the conclusions is O($|D|^2$).
\end{proof}
}

Recall that %$\cS$ is a set of semantics, 
$\cS = \{ \mi{partial~stable, well\mbox{--}founded, regular, L\mbox{--}stable, Kunen, Fitting} \}$
is a set of semantics.
These semantics (and the stable semantics) are preserved by unfolding (with the Kunen semantics requiring the restriction on a rule unfolding itself).
This was established for the well-founded \cite{Seki93,Aravindan} and stable models \cite{unstable,Aravindan},
and in \cite{cdr} for the partial stable models and the L-stable models.
For the Kunen and Fitting semantics it follows the same proof as in \cite{trans}
for the 2-valued Clark completion semantics.
Consequently, Theorem \ref{thm:correct} extends to the semantics in $\cS$:

\addtocounter{theorem}{1}

\begin{theorem}   \label{thm:correct2}
Let $D = (F, R, >)$ be a defeasible theory, and $\alpha$ be an annotation for that theory.
Let $d \in \{ \delta^*, \delta, \partial^*, \partial \}$. 
% \supp_\partial, \supp_{\partial^*}, \supp_\delta, \supp_{\delta^*} \}$.
Suppose $\alpha(R)$ contains only annotations $\free$ and $d$, and there is no fail-expression in $R$. 
Let $S \in \cS$.
Then

\begin{itemize}
\item
$\cM(\alpha(D)) \models_S \mt{defeasibly_d}(q)$ iff 
$\cM_d(D) \models_S \mt{defeasibly}(q)$

\item
$\cM(\alpha(D)) \models_S \neg \mt{defeasibly_d}(q)$ iff 
$\cM_d(D) \models_S \neg \mt{defeasibly}(q)$
\end{itemize}

\noindent
and, if $d \in \{ \delta^*, \delta \}$,
\begin{itemize}
\item
$\cM(\alpha(D)) \models_S \mt{supported_d}(q)$ iff 
$\cM_d(D) \models_S \mt{supported}(q)$
\item
$\cM(\alpha(D)) \models_S \neg \mt{supported_d}(q)$ iff 
$\cM_d(D) \models_S \neg \mt{supported}(q)$
\end{itemize}

More generally,
the S-models of $\cM(\alpha(D))$ restricted to $\mt{defeasibly_d}$ are identical (up to predicate renaming) 
to the S-models of $\cM_d(D)$ restricted to $\mt{defeasibly}$.
\end{theorem}
\skipit{
\begin{proof}
The proof of Theorem \ref{thm:correct} also applies to this theorem,
since unfolding (without self-unfolding) preserves models for all semantics in $\cS$
(see Theorem 3.2 of \cite{cdr}), as does deletion of irrelevant clauses.
\end{proof}
}

\section{Examples}

We present some counterexamples, to show that Figure \ref{fig:inclusion} does not omit any containments
and that all the containments are strict.
For these examples we do not need to use any annotations:
they equally apply to (unannotated) defeasible theories, and we present them in that form.

There are four possible containments we must show do not hold:
$\delta \not\subseteq \partial^*$,
$\supp_\delta \not\subseteq \supp_{\partial^*}$, 
$\partial \not\subseteq \supp_\delta$, 
and
$\delta \not\subseteq \supp_{\partial^*}$.
We have two examples that demonstrate these four points.

\begin{example}
% $\delta \not\subseteq \partial^*$
% $\supp_\delta \not\subseteq \supp_{\partial^*}$

Let the defeasible theory $D$ consist of the rules
\[
\begin{array}{lrcl}
r_1: &           & \Rightarrow & \phantom{\neg} p \\
r_2: &           & \Rightarrow & \neg p \\
r_3: &           & \Rightarrow & \phantom{\neg} q \\
r_4: &  \neg p & \Rightarrow & \neg q \\
r_5: &     q    & \Rightarrow & \phantom{\neg} s \\
r_6: &           & \Rightarrow & \neg s \\
\end{array}
\]
with $r_5 > r_6$.

Rules $r_1$ - $r_4$ are a standard example distinguishing ambiguity blocking and propagating behaviours.
$+\partial^* q$ and $-\delta q$ can be concluded.
Thus, $\delta \not\subseteq \partial^*$.
In addition, we conclude $-\supp_{\partial^*} \neg s$ and $+\supp_\delta \neg s$.
Thus, $\supp_\delta \not\subseteq \supp_{\partial^*}$.

\end{example}

Now we show that $\partial \not\subseteq \supp_\delta$ and $\delta \not\subseteq \supp_{\partial^*}$.
\begin{example}  
% $\partial \not\subseteq \supp_\delta$, 
% $\delta \not\subseteq \supp_{\partial^*}$.

Consider the following defeasible theory $D$:
\[
\begin{array}{lrclclrcl}
r_1: &    p     & \Rightarrow &                 \neg q & ~~~~~~~~~~~&
r_5: &           & \Rightarrow & \phantom{\neg} p \\

r_2: &           & \Rightarrow &\phantom{\neg} q & &
r_6: &           & \Rightarrow & \phantom{\neg} p \\
r_3: &    q     & \Rightarrow &                 \neg s & &
r_7: &           & \Rightarrow & \neg p \\
r_4: &           & \Rightarrow & \phantom{\neg} s & &
r_8: &           & \Rightarrow & \neg p \\
\end{array}
\]
with $r_1 > r_2$, $r_5 > r_7$, $r_6 > r_8$

Then we have $+\delta p$ and ${-}\partial^* p$.
Consequently, we have ${-}\supp_\delta q$ and ${+}\partial^* q$.
Hence, $\partial \not\subseteq \supp_\delta$.
Furthermore, we have ${+}\delta s$ and ${-}\partial^* s$.
Hence $\delta \not\subseteq \supp_{\partial^*}$.
\end{example}

Hence, there are no containments missing from Figure \ref{fig:inclusion}.

% strictness
That the containments in the top row of Figure \ref{fig:inclusion} are strict
was mostly established in \cite{TOCL10}.
The strictness of containments between forms of support follows straightforwardly
from the strictness of containment for the corresponding forms of defeasibility. 
For the remaining containments, consider the following example.

\begin{example}  \label{ex:ABAP}
Consider the following defeasible theory $D$:
\[
\begin{array}{lrl}
r_1:  & p  & \Rightarrow q \\

r_2: &   & \Rightarrow \neg q \\

r_3:  &   & \Rightarrow p \\

r_4: &   & \Rightarrow \neg p \\
\end{array}
\]
We have $-\partial^* p$ (and $-\partial^* \neg p$) but $+\supp_{\delta^*} p$.
%Thus $\partial^*$ and $\supp_{\delta^*}$ differ on $D$.
Consequently, we have $+\partial^* \neg q$ but $-\delta^* \neg q$,
showing that $\delta^* \subset \partial^*$ on $D$.

Note also that we have conclusions ${-}d p$ and  $+\sigma_d p$ for any defeasible proof tag $d$.
Hence $+\partial \subset \sigma_{\partial}$ and $+\partial^* \subset \sigma_{\partial^*}$
\end{example}

Hence all the containments in Figure \ref{fig:inclusion} are strict.

\end{appendix}

\end{document}